\documentclass[aps,twocolumn, superscriptaddress , amsmath, amssymb, showpacs, showkeys,reprint]{revtex4-1}
\usepackage{graphicx}% Include figure files
\usepackage{dcolumn}% Align table columns on decimal point
\usepackage{bm}% bold math
\usepackage{upgreek}% Greak symbols NOT italic
\usepackage{times}
\usepackage{mathrsfs}
\usepackage{multirow}
\usepackage{fancyhdr} % header/footers
\usepackage{hyperref} % hyperlinks
\begin{document}

%\preprint{ABC}
\title{Robust Bloch character at the band edges of hybrid halide perovskites}
\author{O.~Rubel}
\email[]{rubelo@mcmaster.ca}
\affiliation{Department of Materials Science and Engineering, McMaster University, 1280 Main Street West,
Hamilton, Ontario L8S 4L8, Canada}
\affiliation{Thunder Bay Regional Research Institute, 980 Oliver Road, Thunder Bay, Ontario P7B 6V4, Canada}
\author{A.~Bokhanchuk}
\affiliation{Thunder Bay Regional Research Institute, 980 Oliver Road, Thunder Bay, Ontario P7B 6V4, Canada}
\affiliation{Confederation College, 1450 Nakina Dr., Thunder Bay, Ontario P7C4W1, Canada}

\date{\today}

\begin{abstract}
The high-symmetry pseudocubic init cell is often used for modelling the electronic structure of halide perovskites. However, direct comparison of the band structure with more realistic low-symmetry phases is impeded by the zone folding. We utilize a spectral density technique in order to reduce the supercell band structure to a common Bloch basis. This allows us to compare the electronic structure of high- and low-symmetry phases as well as investigate effects of structural and compositional disorder on states near to the band edges that govern transport, dissociation and recombination of optical excitations.
\end{abstract}


\maketitle

%-----------------------------------------------------------------------
%
%                       I N T R O D U C T I O N
%
%-----------------------------------------------------------------------
\section{Introduction}\label{Sec:Introduction}

A new class of organic-inorganic hybrid perovskite materials, such as (CH$_3$NH$_3$)PbI$_3$, has recently emerged as an alternative choice for photovoltaic devices and immediately drew enormous attention \cite{Burschka_N_499_2013,Liu_NP_8_2014,Zhou_S_345_2014,Jeon_NM_13_2014}. (CH$_3$NH$_3$)PbI$_3$ possesses a combination of characteristics that make it attractive for solar cell applications including a favourable band gap of 1.6~eV, efficient optical adsorption, high level of mobility for charge carries of both polarities and exceptionally long carrier lifetime \cite{Stoumpos_IC_52_2013}. Device characteristics, such as the power conversion efficiency, can be further enhanced by using mixed halide perovskites, e.g., (CH$_3$NH$_3$)Pb(Cl$_{x}$I$_{1-x}$)$_3$ \cite{Stranks_S_342_2013,Chen_NC_6_2015}. The demonstrated energy conversion efficiency of nearly 20\%~\cite{Park_JPCL_4_2013,Zhou_S_345_2014} combined with low-temperature solution processing technology hold a promise to lead the way towards low-cost solar energy. 

The crystal structure of (CH$_3$NH$_3$)PbI$_3$ exhibits several polymorphs. At low temperatures, an orthorhombic $\gamma$-phase prevails up to 161~K \cite{Baikie_JMCA_1_2013}. A tetragonal $\beta$ structure is stable in the intermediate temperature range of 161--330~K. The high-temperature cubic $\alpha$-phase exists above 330~K \cite{Baikie_JMCA_1_2013}. Electronic structure calculations performed for the cubic phase reveal a direct band gap of approximately 1.3~eV at the R$(1/2,1/2,1/2)$ point of the Brillouin zone \cite{Chang_JKPS_44_2004,Baikie_JMCA_1_2013,Brivio_PRB_89_2014}. The direct gap moves to $\Gamma(0,0,0)$ as the cubic structure transforms into the tetragonal phase \cite{Menendez-Proupin_PRB_90_2014,Umari_SR_4_2014}, which has a more complex unit cell with larger volume. The transformation is accompanied by widening of the band gap, which reaches the value of 1.6~eV as determined by photoluminescence measurements \cite{Yamada_APE_7_2014} and confirmed by \textit{ab initio} calculations using hybrid functionals and quasiparticle GW approaches \cite{Menendez-Proupin_PRB_90_2014,Umari_SR_4_2014}.

Further analysis of the band structure evolution in the course of the transition from high- to low-symmetry structure is hindered by the Brillouin zone folding resulted from the increased size of the unit cell for $\beta$-phase. Here we show that a direct comparison of the electronic structure between $\alpha$ and $\beta$ phases of (CH$_3$NH$_3$)PbI$_3$ can be achieved by projecting the band structure of $\beta$-phase onto the Brillouin zone of the cubic structure, which represents an idealized version of low-symmetry phases. This analysis can facilitate answering the following questions: (i) to which extent does the Bloch character of band edges in the tetragonal phase inherit the R-character of the cubic phase, and (ii) what is the effect of dynamic disorder in the orientation of organic groups or substitutional alloying, e.g., in the mixed halide (CH$_3$NH$_3$)Pb(Cl$_{x}$I$_{1-x}$)$_3$ perovskite, on the Bloch states at the band edges? An uncertainty in the Bloch character of states near to the band edges in semiconductors can be used to reveal  deterioration of charge transport characteristics due to carrier localization \cite{Wang_PRL_80_1998,Popescu_PRL_104_2010,Rubel_PRB_90_2014}, which can adversely affects the performance of solar cells.

%-----------------------------------------------------------------------
%
%                       M E T H O D
%
%-----------------------------------------------------------------------
\section{Computational details}\label{Sec:Method}

The first-principles calculations were carried out using density functional theory (DFT) \cite{Kohn_PR_140_1965}. \texttt{ABINIT} package \cite{Gonze_CMS_25_2002,Gonze_ZK_220_2005} and projector augmented-wave method were used \cite{Torrent_CMS_42_2008} with pseudopotentials adopted from a \texttt{JTH} library (version 0.2) \cite{Jollet_CPC_185_2014}. \citet{Perdew_PRL_77_1996} construction of the generalized gradient approximation (GGA-PBE) was employed for the exchange correlation functional. Completeness of the plane wave basis set was controlled though the cut-off energy, which was set at 15~Ha (408~eV) as determined from convergence studies.

\begin{figure}
	\includegraphics[width=0.3\textwidth]{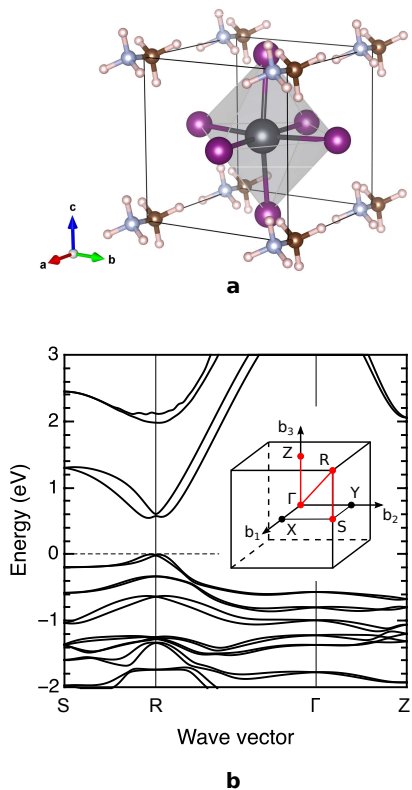}\\
	\caption{Pseudocubic structure of (CH$_3$NH$_3$)PbI$_3$ (a) and the corresponding relativistic band structure (b) along high-symmetry path in the Brillouin zone shown in the inset.}\label{Fig:A}
\end{figure}

The lattice constants of (CH$_3$NH$_3$)PbI$_3$ in pseudocubic phase (Fig.~\ref{Fig:A}(a)) was optimized self-consistently by simultaneous minimization of stresses and Hellmann-Feynman forces acting on atoms below 1~mHa/Bohr. The Brillouin zone was sampled using $8\times8\times8$ \citet{Monkhorst_PRB_13_1976} k-mesh. The calculated lattice constant of 6.47~{\AA} is slightly overestimated as compared to the experimental value of 6.28~{\AA} \cite{Baikie_JMCA_1_2013}. This result is consistent with typical trends for modelling of solids with GGA-PBE exchange-correlation energy functional \cite{Haas_PRB_79_2009}.

\begin{figure*}
    \includegraphics[width=0.9\textwidth]{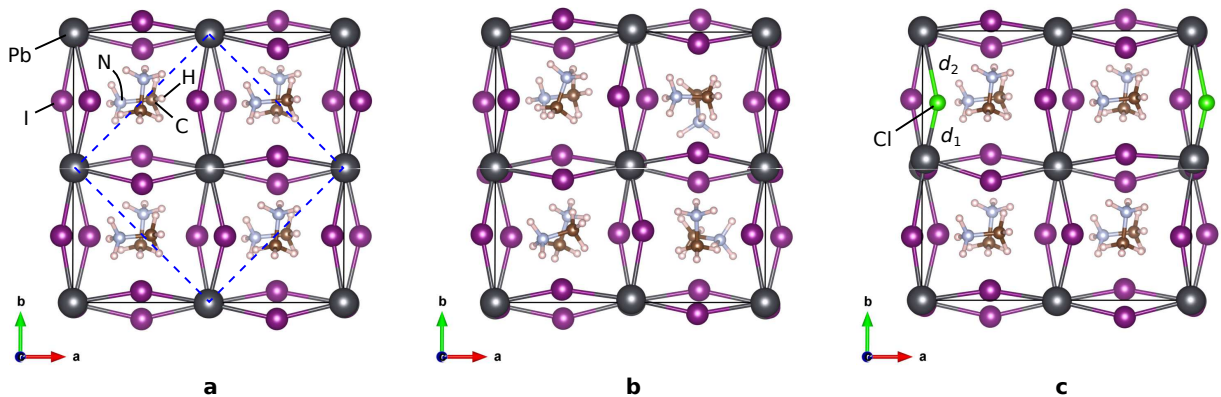}\\
	\caption{Structure of $\beta$-(CH$_3$NH$_3$)PbI$_3$ (a), $\beta$-(CH$_3$NH$_3$)PbI$_3$ with a disorder in orientation of methylammonium molecules (b) and $\beta$-(CH$_3$NH$_3$)Pb(Cl$_{x}$I$_{1-x}$)$_3$ with $x\approx0.014$ (c). The conventional tetragonal unit cell is outlined by a dashed contour on the panel (a).}\label{Fig:B}
\end{figure*}

The structure of tetragonal phase was constructed by repeating the cubic cell with the multiplicity of $2\times2\times2$ (Fig.~\ref{Fig:B}(a)). The octahedral tilting and orientation of methylammonium molecules were initially set based on previously reported structures of $\beta$-(CH$_3$NH$_3$)PbI$_3$ resolved experimentally and theoretically \cite{Dang_CEC_17_2015,Yin_JPCC_119_2015}. Then the supercell was fully relaxed. Calculated lattice parameters for tetragonal phase $a_0=9.02$~{\AA} and $c_0=13.01$~{\AA} are within 4\% deviation from the corresponding experimental values at room temperature $a_0=8.85$~{\AA} and $c_0=12.52$~{\AA} \cite{Baikie_JMCA_1_2013}. It should be mentioned that our tetragonal supercell differs from a conventional tetragonal unit cell by 45$^{\circ}$ rotation. The construction of the supercell by translation of the parent (cubic) structure along its lattice vectors allows us later to unfold the band structure and present the electron wave vectors in terms of the cubic reciprocal lattice.

Unfolding of the band structure of supercells was performed by calculating the Bloch spectral weight using \texttt{fold2Bloch} package \cite{Rubel_PRB_90_2014} implemented in \texttt{ABINIT}.

\texttt{VESTA}~3 package \cite{Momma_JAC_44_2011} was used for atomic structure visualization.

%-----------------------------------------------------------------------
%
%                       R E S U L T S
%
%-----------------------------------------------------------------------
\section{Results and discussion}\label{Sec:Results}

We begin the analysis of the electronic structure of (CH$_3$NH$_3$)PbI$_3$ with the high-symmetry pseudocubic structure. The calculated relativistic band structure is shown in Fig.~\ref{Fig:A}(b). The band gap is direct near R$(1/2,1/2,1/2)$ point of the Brillouin zone. The magnitude of the band gap is underestimated by approximately 1~eV in comparison to the experimental value of 1.6~eV \cite{Yamada_APE_7_2014}, which is a typical result for DFT calculations of (CH$_3$NH$_3$)PbI$_3$ when the bear GGA-PBE exchange-correlation functional is combined with spin-orbit coupling. The agreement can be further improved by using a modified Becke-Johnson potential \cite{Tran_PRL_102_2009}, however it is not required here due to the comparative nature of our study. A Rashba-like breaking of the spin degeneracy \cite{Brivio_PRB_89_2014} is observed at the band edges (particularly strong in the conduction band). \citet{Zheng_ArXiv_1505.04212_2015} reported a different chirality of the spin texture at the top valence band and the bottom conduction bands that results in the long carrier lifetime due to constrains imposed by the spin section rules.

In order to elucidate origin of the Rashba splitting, the band structure calculation was performed for CsPbI$_3$ in the pseudocubic structure (not shown here). Atomic position and lattice perimeters were kept at the equilibrium values of (CH$_3$NH$_3$)PbI$_3$, except for the methylammonium that was substituted by Cs. The band gap and the Rashba splitting near to the band edges of the pseudocubic CsPbI$_3$ was found to be identical to those in (CH$_3$NH$_3$)PbI$_3$. Furthermore, the splitting is fully suppressed in the centrosymmetric cubic CsPbI$_3$. This result suggests that precursors for the Rashba splitting are the strong spin-orbit interaction and distortions in the Pb-centred octahedron caused by the dipolar nature of CH$_3$NH$_3^+$ ions \cite{Zheng_ArXiv_1505.04212_2015}. A significant reduction of the Rashba splitting observed in (NH$_4$)PbI$_3$ perovskite structures \cite{Brivio_PRB_89_2014} confirms validity of this argument.

The cubic structure is attractive for modelling due to its simplicity. However, we should keep in mind that the pseudocubic structure is an idealization; the room temperature properties are governed by the tetragonal phase shown in Fig.~\ref{Fig:B}(a). Distinctive features of the tetragonal structure are the presence of an octahedral tilting and CH$_3$NH$_3^+$ cation rotation. This leads to the question as to which extend the discussion of the electronic structure of the pseudocubic phase is applicable to the tetragonal structure? In order to enable direct comparison between two electronic structures, the band structure of low-symmetry tetragonal phase need to be mapped onto the pseudocubic Brillouin zone. This can be achieved by ``unfolding" the supercell band structure. Since the tetragonal structure comprises of $2^3$ pseudocubic cells, each eigenstate $E_\mathbf{K}$ in the supercell gives rise to eight k-points in the pseudocubic Brillouin zone. For instance, $\Gamma(0,0,0)$ unfolds into a set of eight k-points, three of which are obtained by permutations of $(1/2,0,0)$, another three additional k-points are permutations of $(1/2,1/2,0)$, one k-point is at $(1/2,1/2,1/2)$ and the $\Gamma$-point self.

It is evident that $\Gamma$-point in tetragonal structure is coupled to R-point in the pseudocubic structure. However, it is not possible to say \textit{a priori} what will be the strength of R-representation. Each unfolded k-point is assigned its own Bloch spectral weight $w_\mathbf{k}$. The whole set satisfies normalization $\sum_\mathbf{k} w_\mathbf{k}=1$. The spectral weight us used as an augmentation for the band structure plot $E(\mathbf{k},w_\mathbf{k})$ and allows to filter out spurious bands. The spectral weight is computed by projecting the supercell wavefunction $\Psi_\mathbf{K}$ onto a set of plane waves that are compatible with the wave vector $\mathbf{k}$ and  reciprocal lattice vectors of the pseudocubic Brillouin zone. Details of this procedure are widely discussed in the literature \cite{Wang_PRL_80_1998,Popescu_PRL_104_2010,Popescu_PRB_85_2012,Allen_PRB_87_2013,Medeiros_PRB_89_2014,Rubel_PRB_90_2014}.

\begin{figure*}
    \includegraphics[width=0.95\textwidth]{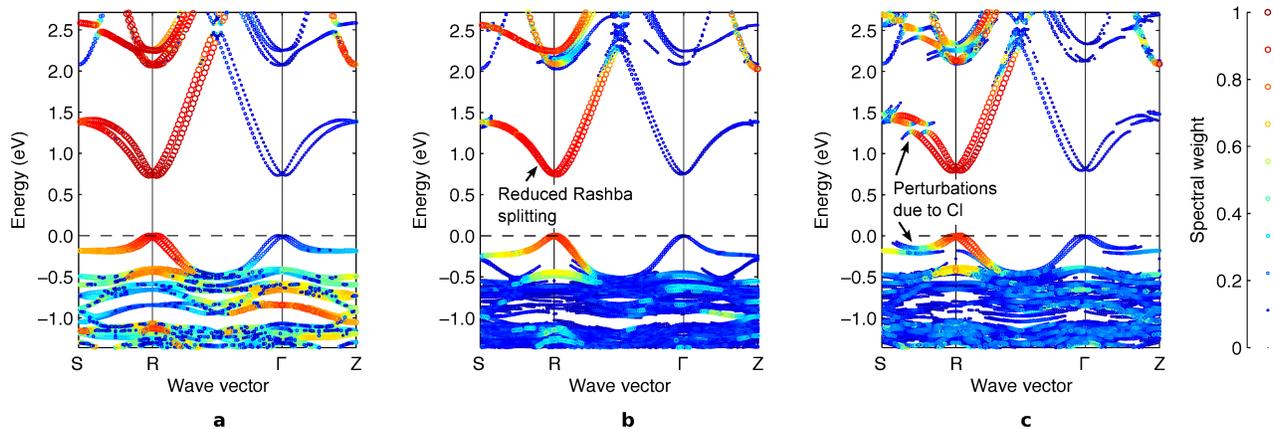}\\
	\caption{Effective relativistic band structure of $\beta$-(CH$_3$NH$_3$)PbI$_3$ (a), $\beta$-(CH$_3$NH$_3$)PbI$_3$ with a disorder in orientation of methylammonium molecules (b) and $\beta$-(CH$_3$NH$_3$)PbI$_{3(1-x)}$Cl$_{3x}$ with $x\approx0.014$ (c). The size and colour of the markers correspond to the spectral weight of supercell eigenstates projected to the Bloch states in the pseudocubic Brillouin zone (see inset in Fig.~\ref{Fig:A}(b)).}\label{Fig:C}
\end{figure*}

The effective band structure for tetragonal phase is shown in Fig.~\ref{Fig:C}(a). The band gap increases by approximately 0.1~eV when compared to the pseudocubic structure. The conduction and valence band edges of tetragonal phase retain 90 and 84\% of the pseudocubic R-character, respectively. The $\Gamma$-character of the corresponding states is only 10\% at most. The tetragonal phase exhibits a similar magnitude of the Rashba splitting as in the pseudocubic structure (Fig.~\ref{Fig:A}(b)).

Now as the basis for comparison of the electronic structure for high- and low-symmetry phases is established, we can proceed with disordered structures. A low energy barrier for rotation of the CH$_3$NH$_3^+$ ion and occurrence of multiple local minima  \cite{Zhou_PCCP_unknown_2015} leads to discretized reorientation dynamics of the C--N axis in (CH$_3$NH$_3$)PbI$_3$ at the room temperature \cite{Leguy_NC_6_2015}. The model for disordered orientation of methylammonium molecules was generated from $\beta$-(CH$_3$NH$_3$)PbI$_3$ structure (Fig.~\ref{Fig:B}(a)) by flipping the molecules 180$^\circ$ randomly with subsequent relaxation of atomic coordinates. The obtained structure is shown in Fig.~\ref{Fig:B}(b). It should be noted that the total energy of the disordered structure was even lower in comparison to the unperturbed tetragonal phase.

The band structure of a disordered tetragonal phase is presented in Fig.~\ref{Fig:C}(b). The disorder does not lead to significant changes in the band gap. The band edges retain their R-character at the level of $83-88$\%, which is almost identical to that in the unperturbed tetragonal phase. Thus we can conclude that the orientational disorder of organic groups does not introduce localization effects at the band edges and does not hamper transport of charge carriers. This interpretation is based on the relation between  the ambiguity of the Bloch character in reciprocal space and localization in the real space through the fundamental uncertainty principle. The only noticeable effect of the orientational thermal disorder is the reduced Rashba splitting, which can make the spin selection rule, which promotes dissociation of optical excitations, less efficient.

Finally, we turn the discussion to a compositional disorder in mixed halide perovskite structures. Even though the Cl incorporation in an iodide-based structure is relatively low (below 3--4\%) \cite{Colella_CM_25_2013}, the recombination time and diffusion length of optical excitations are prolonged by 1--2 orders of magnitude \cite{Stranks_S_342_2013,Wehrenfennig_AM_26_2014}. The origin for a superior performance of (CH$_3$NH$_3$)Pb(Cl$_{x}$I$_{1-x}$)$_3$ perovskite thin film solar cells remains an open question. The plausible scenarios include Cl$^-$ ions involved in the following functions:  facilitation the release of excess CH$_3$NH$_3^+$ during annealing \cite{Yu_AFM_24_2014}, passivation of grain boundaries and the interface with the substrate \cite{Unger_CM_26_2014,Mosconi_JPCL_5_2014}, creation of an in-homogeneous built-in electric field profile similar to p-i-n junction \cite{Edri_NC_5_2014,Starr_EES_8_2015}, and reduction of the carrier effective mass in structures with high Cl content \cite{Li_JAP_117_2015}.

In order to model a dilute substitutional (CH$_3$NH$_3$)Pb(Cl$_{x}$I$_{1-x}$)$_3$ alloy, one of 24 I-atoms in the supercell was replaced by Cl  that resulted in the effective composition of $x\approx0.014$. The structural relaxation in the vicinity of chlorine reveals asymmetric Pb-Cl bonds $d_1<d_2$ (see Fig.~\ref{Fig:B}(c)). The Pb-Cl bond lengths of 2.75 and 3.48~{\AA} deviate significantly from the Pb-I equilibrium bond length of 3.25~{\AA}. The shorter Pb-Cl bond length agrees with the sum of Pb$^{4+}$ and Cl$^-$ ionic radii, 0.94 and 1.81~{\AA} respectively \cite{Shannon_ACA_32_1976}. The asymmetric configuration could be an efficient way to mediate local strain due to the large size mismatch between cations and can be a driving force for segregation of Cl at interfaces.

The effective band structure of (CH$_3$NH$_3$)Pb(Cl$_{0.014}$I$_{0.986}$)$_3$ is shown in Fig.~\ref{Fig:C}(c). Incorporation of chlorine in dilute quantities does not change the band gap, which is in accord with observations made by \citet{Colella_CM_25_2013}. The Rashba splitting and the Bloch character of the valence and conduction band edges remain intact with the undopped (CH$_3$NH$_3$)PbI$_3$. Perturbations due to Cl are observes for states \textit{within} the conduction and valence bands as pointed out by arrows in Fig.~\ref{Fig:C}(c). Since the access energy required to reach perturbed states is greater than the thermal energy, those perturbations in the band structure will unlikely have an adverse effect on transport coefficients of the mixed halide perovskites.

%Band Eg increases with volume \cite{Feng_JPCC_118_2014}, interesting

%-----------------------------------------------------------------------
%
%                       C O N C L U S I O N S
%
%-----------------------------------------------------------------------
\section{Conclusions}\label{Sec:Conclusions}

The relativistic band structure of low-symmetry $\beta$-(CH$_3$NH$_3$)PbI$_3$ tetragonal phase is mapped onto the pseudocubic Brillouin zone using an unfolding technique. In spite of significant differences in the structure (the presence of an octahedral tilting and CH$_3$NH$_3^+$ cation rotation), the band structures of both phases share many similarities. A strong coupling is observed between $\Gamma$-point in tetragonal structure and R-point in the pseudocubic structure. The conduction and valence band edges of tetragonal phase retain 80--90\% of the pseudocubic R-character. The $\Gamma$-character of the corresponding states is very weak in contrast to the common expectation. Both pseudocubic and tetragonal structures exhibits a similar magnitude of the Rashba splitting, which plays an important role in the long carrier lifetime due to constrains imposed by the spin section rules. At room temperature, the Rashba splitting can be reduced due to an orientational disorder of CH$_3$NH$_3^+$ cations. At the same time, the disorder neither affects the band gap nor reduces the R-character of the band edges and thus does not hamper transport of charge carriers.

Mixed halide (CH$_3$NH$_3$)Pb(Cl$_{x}$I$_{1-x}$)$_3$ structures with dilute concentration of chlorine reveal existence of asymmetric Pb-Cl bonds as a way to mediate local strain due to the large size mismatch between cations that can also be a reason for segregation of Cl at interfaces. The band gap, Rashba splitting and the Bloch character of the valence and conduction band edges in perovskite structures with dilute substitutional chlorine remain intact with the undopped (CH$_3$NH$_3$)PbI$_3$. Perturbations due to the incorporated chlorine are observes for states located energetically far from the band edges. This rules out the possibility of adverse effect of chlorine substitutional incorporation on transport coefficients of the mixed halide perovskites.

%-----------------------------------------------------------------------
%
%                       A C K N O W L E D G E M E N T
%
%-----------------------------------------------------------------------
\begin{acknowledgments}
The work was supported by the Natural Sciences and Engineering Research Council of Canada under the Discovery Grant Program RGPIN-2015-04518. Computational resources were provided by High Performance Computing Centre at Lakehead University and Compute Canada.
\end{acknowledgments}

%                       B I B L I O G R A P H Y
%
%-----------------------------------------------------------------------
%\clearpage
%\bibliography{../bibliography}

\end{document}